\documentclass[twocolumn]{aastex701}
\usepackage{epstopdf}
\usepackage{amsmath}
\usepackage{amssymb}
\usepackage{natbib}
\usepackage{listings}
\usepackage{graphicx}
\newcommand{\WAsp}         {{\tt WAsp}}
\newcommand{\Asp}         {{\tt Asp}}
\newcommand{\transpose}         {{\rm T}}

\def\tens#1{\ensuremath{\mathsf{#1}}}
\usepackage{txfonts}
\usepackage{framed}
\usepackage{bm} 
\def\BF#1{%
  \ifmmode
    \bm{#1}
  \else
    \textbf{#1}
  \fi
}
\def\BF#1{{#1}}

\begin{document}

\title{WAsp: The Wideband (W) Adaptive-Scale Pixel (Asp) Deconvolution
  Algorithm for Interferometric Imaging}

\author{M.~Hsieh}
\affiliation{National Radio Astronomy Observatory, 1011 Lopezville Road, Socorro, NM 87801, USA}
\email[show]{mhsieh@nrao.edu}

\author{S.~Bhatnagar}\affiliation{National Radio Astronomy Observatory, 1011 Lopezville Road, Socorro, NM 87801, USA}
\email[show]{sbhatnag@nrao.edu}

\author{U.~Rau}\affiliation{National Radio Astronomy Observatory, 1011 Lopezville Road, Socorro, NM 87801, USA}
\email[show]{rurvashi@nrao.edu}
\submitjournal{The Astronomical Journal (AJ)}
\accepted{April 23, 2026}
\correspondingauthor{S. Bhatnagar}

\begin{abstract}
This paper introduces the Wide-band Asp-Clean (\texttt{WAsp}) algorithm, a
novel scale-sensitive image reconstruction method tailored for
wide-band imaging applications. This algorithm is particularly
beneficial for thermal noise-limited imaging with aperture synthesis
telescopes, where joint spatio-frequency modeling of the sky
brightness distribution is critical. The \WAsp\ algorithm replaces the use
of the MS-Clean algorithm in the MS-MFS algorithm \citep{MSMFS} with
the {\tt Asp} algorithm \citep{Asp_Clean}, which itself has been improved for both
imaging and runtime performance.  With the high sensitivity of current
and next-generation telescopes, spatio-frequency modeling in a
scale-sensitive basis becomes crucial for ensuring that residuals
align with the noise model across the frequency band. Although
existing wide-band scale-sensitive algorithms have demonstrated
superior performance over scale-insensitive counterparts, they often
suffer from well-documented deficiencies, leading to significant
wide-scale residuals in Stokes-I at low levels and consequently
significant relative errors in spectral index maps.

The \WAsp\ algorithm addresses these limitations while maintaining 
computational efficiency.  The implementation can be configured to 
support narrow-band and wide-band scale-sensitive imaging, 
spectral-cube imaging applications and joint single-dish and 
interferometer imaging. To demonstrate improved imaging
performance, we show comparison with existing algorithms via carefully
developed simulations for stress-testing the algorithms. We also
present results from its application to real-world wide-band data,
underscoring its effectiveness in practical imaging scenarios.
\end{abstract}

\section{Introduction}
Aperture synthesis telescopes are indirect imaging devices which
collect data in the Fourier domain.  The sampling of the data domain
is also typically incomplete.  Mapping the sky brightness distribution
with such telescopes therefore require iterative algorithms for
transforming the data to the image domain, and reconstruction of the
sky brightness distribution using modeling algorithms.  In the
numerical optimization terminology, each iteration involves computing
the derivative of the cost function with respect to the model image,
and using the derivative to update the image model such that the
residuals in each iteration progressively become noise-like.  These map
on to the {\tt UpdateDir} and {\tt ModelUpdate} components of the
algorithm architectural description in \cite{AlgoArch}.  In the
traditional radio astronomy parlance, these steps are referred to as
the ``major cycle'' and the ``minor cycle'', respectively.  We use
{\tt UpdateDir} and ``major cycle'', and {\tt ModelUpdate} and ``minor
cycle'' interchangeably throughout the text.

Using the terminology from \citep{Asp_Clean}, the measurement equation
describing an interferometer can be written as:
\begin{equation}
\label{ME}
 \vec{V} = \tens{A} \vec{I}^o + \tens{A} \vec{N}
\end{equation}
where $\vec{V}$ is the the measured visibility, $\vec{N}$ is the 
independent random noise vector, and $\vec{I}^o$ is the
true image.  The measurement matrix $\tens{A}$ represents the linear
transform from the image to the visibility domain.

The primary goal of algorithms for imaging can be
mathematically described as a search for the model $\vec{I}^M$ such
that $\vec{V} - \tens{A} \vec{I}^M$ is statistically consistent with
$\tens{A}\vec{N}$.  In practice, $\tens{A}$ in general is a singular
non-square matrix requiring non-linear methods (linear methods to
invert Eq.~\ref{ME} are fundamentally excluded).  In the radio
astronomy parlance, $\tens{A}^\transpose \vec{V}$ is the dirty image
($\vec{I}^D$), $\tens{A}^\transpose \tens{A}=\tens{B}$ is the beam matrix
(the super-script $\rm T$ implies a transpose) and $\tens{B}\vec{I}^o$
represents the convolution of the image with the PSF.  Note that the
noise vector in the image domain ($\tens{A}^\transpose \tens{A} \vec{N}$)
is also convolved with the PSF leading to correlated noise between
adjacent pixels.

The model $\vec{I}^M$ is parameterized, and the nature of the search 
space depends on how compatible the model is with the instrument sampling 
function and sky structure (see \citep{Asp_Clean} for a more detailed 
overview).  Briefly, scale {\it insensitive} algorithms 
(e.g. H\"{o}gbom-CLEAN \citep{Hogbom_Clean}, MEM \citep{MEM_ARAA} and 
their variants) parameterized the image as a
collection of independent delta-functions of amplitude $F$ located at
$(x_k,y_k)$ as:
\begin{equation}
\label{MODIMG}
\vec{I}^M=\sum_k F_k\delta(x-x_k,y-y_k)
\end{equation}
The algorithms solve for $F_k$ within the search domain as independent
parameters.  This class of algorithms perform poorly in modeling large
scale emission and leave residuals.  The detailed structure of these
residuals depends on the regularization terms used in different
algorithms.  For wide-band imaging, such residuals in Stokes-I images
lead to significantly higher errors in modeling of the frequency
dependence of the brightness distribution (e.g. in the spectral index
maps).  Scale-sensitive algorithms (like MS-Clean, \Asp-Clean and
their variants) on the other hand parameterize the sky brightness
distribution as
\begin{equation}
\vec{I}^M = \sum_k P_k(\vec{p})
 \label{Eq:IMOD}
\end{equation}
where $P_k$ is the parameterized scale-sensitive component, which we
refer to as the {\it Aspen}. The $\vec{p}$ is the vector of parameters
(such as $\{Amplitude, Location, Scale\}$), which are solved to model
both compact and large scale emission.  The Aspen $P(\vec{p})$ can be
any appropriate functional form, including a symmetric Gaussian
\citep{Asp_Clean}. Algorithms in this class are inherently
scale-sensitive and perform significantly better than {\it
  scale-insensitive} algorithms in modeling both compact and large
scale emission.  In general these algorithms reconstruct the Stokes-I
emission significantly better, and therefore also model the
spatio-frequency structure significantly better.  Although these
algorithms formally incur a higher computational cost per iteration,
they typically exhibit a faster rate of convergence. This improved
convergence behavior partially offsets their increased per-iteration
expense. In many practical applications, the dominant computational
cost arises from the evaluation of the derivative (the {\tt UpdateDir}
step (referred to as the “major cycle” in radio astronomy literature).
In such cases, the reduced number of required iterations can lead to a
lower overall runtime (see \cite{Asp_Clean} for a more detailed
overview).

Although formally
these algorithms have higher runtime cost, their rate of convergence
is typically higher which partly compensates for their overall runtime
cost.  For many use-cases the computing cost of calculating the
{\tt UpdateDir} step dominates the overall cost.  For
those cases, scale-sensitive algorithms may actually reduce the
overall runtime cost due to faster rate of convergence (see
\cite{Asp_Clean} for a more detailed overview).

Wide-band image reconstruction is a two-step process.  First, the raw
data needs to be transformed to the image domain in an basis where the
spatio-frequency modeling of the sky brightness distribution can
expressed in the most compact manner.  The primary algorithm for this
step is the Multi-Term Multi-Frequency Synthesis (MT-MFS) algorithm 
which models brightness along the
frequency axis as multiple Taylor-coefficient images.  The 2D
distribution in each of these coefficient images models the spatial
structure.  The second step in the process is to apply one of the
scale-sensitive modeling algorithms, like MS-Clean
\citep{MS-Clean}.  This combination of the MT-MFS
algorithm for frequency modeling and MS-Clean for spatio modeling to
achieve joint spatio-frequency modeling is referred to as the 
Multi-scale Multi-Term Multi-Frequency Synthesis (MS-MFS) \citep{MSMFS} 
algorithm. MS-Clean utilizes a fixed set of functions with predetermined 
scales. At each iteration, an optimal scale is selected from this set and removed from 
the image. Since this set includes both large and small scales, the algorithm 
effectively reconstructs both large and small-scale emissions, leaving residuals at 
lower levels and smaller scales compared to scale-insensitive algorithms. As a result, 
the MS-MFS algorithm achieves a spatio-frequency reconstruction in a scale-sensitive 
basis.

Conceptually, the spectral index maps can be understood as a
difference between Stokes-I maps at different frequencies.  Small
relative errors (residuals) in these Stokes-I maps therefore translate
to much higher equivalent relative errors in the spectral index maps.
So, while the MS-MFS algorithm is a significant improvement over
scale-insensitive algorithms, especially in reconstructing a
combination of compact and diffused emission, the numerical
performance of existing algorithms is insufficient for thermal
noise-limited wide-band imaging, specially in the presence of large
scale emission.  A variant of this approach is an implementation with
scale-dependent masking for runtime optimization \cite{WSCLEAN}.  The
{\tt Resolve} algorithm \citep{ResolveClean} on the other hand is a
MEM-class algorithm formulated in the standard Bayesian inference
framework including algebraic regularization terms to model the sky
brightness and the spectral index simultaneously.  \BF{
In practice, the imaging performance of existing algorithms, not all
of which have demonstrated wide-band modeling capabilities, is
comparable to each other (also see \cite{Zhang_2020, Zhang_2021,
  Mueller-ConvexOpt} for more comparative studies).}

The Wide-band Asp-Clean (\WAsp) algorithm presented here is a
significant improvement over the MS-Clean used in the MS-MFS
algorithmic framework.  It therefore improves the reconstruction of
Stokes-I image, particularly for low-level, large-scale emissions, and
consequently also significantly improving the accuracy of spectral
index maps as well.  Below we describe the algorithm theoretically and
numerically and how it is used for wide-band simultaneous
spatio-frequency image reconstruction, and show results of the
application of \WAsp\ to real wide-band data. As previously noted, the
\WAsp\ algorithm uses the algorithmic framework of the MT-MFS
algorithm to construct Taylor-term images, while incorporating a
modified \Asp-Clean algorithm for modeling in lieu of the MS-Clean
algorithm. A concise review of both the MT-MFS and \Asp-Clean
algorithms is provided in Section~\ref{Sec:background}.

\section{Related Work}
\label{Sec:background}

\subsection{Multi-Scale Multi-Frequency Deconvolution}
\label{Sec:mtmfs}
The MS-MFS algorithm models the wide-band sky brightness 
distribution as a linear combination of spatial and 
spectral basis functions, and performs image reconstruction 
by combining a linear-least-squares approach with iterative 
chi-square minimization. The output of the MS-MFS algorithm is a 
set of $N_t$ Taylor-series coefficient images at $N_s$ different spatial 
scales that represent the spectral structure of the sky brightness 
distribution. The iterative process of MS-MFS that solves the 
normal equations is described below.

\begin{enumerate}
\item  For iteration $i$, the principal solution is computed for all 
  pixels, separately for all scales $s$, resulting in $N_s$ sets of 
  $N_t$ Taylor-coefficient images, referred to as the $TT_t$ images
  where the subscript $t$ indicates the index of the coefficient.
\item Identify the best multi-term flux component over all scales. 
  The result of this step is a set of $N_t$ model images, each 
  containing one $\delta$-function that marks the location of the 
  center of a flux component of shape $I_{p,i}^{shp}$ ($p$ represents 
  the scale of $p,i$ the chosen component, out of all possible values 
  of $s$). The amplitudes of these $N_t$ $\delta$-functions are the 
  Taylor coefficients that model the spectrum of the integrated flux of 
  this component.
  \begin{itemize}
        \item The MS-MFS algorithm applies the MS-Clean algorithm 
          \citep{MS-Clean} in this step.
  \end{itemize}
\item Update the RHS residual images by evaluating and subtracting out
  the entire LHS of the normal equations.
\item Repeat from Step 2 until the minor-cycle flux limit is reached.
\end{enumerate}

\subsection{\Asp-Clean}
\label{Sec:asp}
The original \Asp-Clean algorithm is described in detail in
\cite{Asp_Clean}.  Unlike MS-Clean, \Asp-Clean does not need a
user-provided list of scales but dynamically determines optimal
scales. To accomplish this, two sets of Aspen are maintained -- the 
active set and the permanent set. The active set always keeps the
current optimal scale(s), while the permanent set keeps all the
optimized Aspen found in all previous and the current iterations.  In
the Step 2 (c) below, it evaluates all Aspen in the permanent set onto
an empty model image, and always subtracts that full model image
convolved with the PSF from the original dirty image. That is,
\begin{eqnarray}
  \vec{I}^R &=& \tens{A}^\transpose\vec{V} -  \tens{A}^\transpose\tens{A}\vec{I}^M \nonumber \\
            &=& \vec{I}^D - \tens{B}\vec{I}^M
  \label{Eq:IRES}
\end{eqnarray}
\BF{where $\vec{I}^M$, given by Eq.~\ref{Eq:IMOD}, is constructed of
  the Aspen in the permanent set consisting of all Aspen born in the
  current and all previous iterations,} and $\vec{I}^R$ is the
residual image. The algorithm is composed of the following steps:
\begin{enumerate}
\item Initialize the algorithm by first defining an initial set of
  Aspen with scales as multiples of $W$, the half-width at half-maximum
  of the Gaussian fitted to the main lobe of the PSF.  E.g. the set
  $\left\{W, 2W, 4W, 8W\right\}$.
\item Set $\vec{I}^R = \vec{I}^D$. In each iteration $k$
  \begin{enumerate}
    \item First, smooth $\vec{I}^R$ with the fixed set of scales.
      This gives a set of smoothed residual images.
      \label{Step-1}
    \item Search for the global peak ($F$) among the set of smoothed
      residual images.  Add to the active set an Aspen, $P_k(\vec{p})$, 
      of amplitude $F_k$, scale corresponding the smoothed residual
      image with the global peak, and coordinates of the global peak
      as the location.
      \label{Step-2}
    \item Optimize the set of Aspen(s) in the active set by minimizing
      the objective function, $\vec{I}^R = \vec{I}^D -
      \sum_k\tens{B}P_k(\vec{p})$, where $I^R$ is the {\it current}
      residual image.
    \item Update $\vec{I}^M$ (Eq.~\ref{Eq:IMOD}) and compute
      $\vec{I}^R$ (Eq.~\ref{Eq:IRES}).
    \item Go to \ref{Step-1} unless the termination criteria is met or
      the residuals are noise-like.
  \end{enumerate}
\end{enumerate}

\section{The \WAsp\ Image Reconstruction Algorithm}
\label{Sec:wasp}
In this section, we present the \WAsp\ algorithm for scale-sensitive
wide-band imaging.  The algorithm can be also configured for
narrow-band imaging, making it suitable for spectral cube applications
where it can be applied to individual narrow-band channel images. In
Sec.~\ref{Sec:diffasp} we discuss improvements in the \WAsp\ algorithm
to mitigate some of the runtime inefficiencies of the \Asp-Clean
algorithm, without degrading the imaging performance.
Section~\ref{Sec:diffmtmfs} describes how the multi-term
multi-frequency imaging part of the MS-MFS algorithm is utilized, with
the MS-Clean algorithm replaced by the improved \Asp-Clean
algorithm. Further implementation details are provided in
Appendix~\ref{Appendix:A}.

\subsection{Relation to the Original \Asp-Clean}
\label{Sec:diffasp}
The original \Asp-Clean algorithm gives a significantly better imaging
performance compared to MS-Clean, but Steps~\ref{Step-1} and
\ref{Step-2} in Section~\ref{Sec:asp} become inefficient for complex
images where the total number of Aspen can be large.  To mitigate
this, \cite{Asp_Clean} describe a heuristic approach to speed up the
computation that removes Aspen which have significant effect to begin
with but their modifications are insignificant in the later cycles.

Here, we improved the original approach described in ~\ref{Step-1} for
defining the initial scale sizes to prevent \WAsp\ from picking a
large scale that has no constraints from the data
(Sec.~\ref{Sec:InitScales}). We also developed a new approach that
simplified the algorithm (see Section~\ref{Sec:LatestAspen}) and
reduced the runtime further without degrading the imaging performance.
Finally, we developed a modification of the \cite{FusedAsp} fused
deconvolution algorithm that combines the scale-insensitive H\"{o}gbom
CLEAN algorithm and the \WAsp\ to further improve the imaging
performance and its computational efficiency (Sec.~\ref{Sec:Fused}).

\subsubsection{Define Initial Scale Sizes}
\label{Sec:InitScales}
 The \Asp-Clean defines the initial scale sizes to be 0, $W$, $2W$,
 $4W$, and $8W$. This initial guess consistently reduces convergence 
 time, and has been determined to be robust for most (but not all) 
 cases.  When it fails, it is usually due to the
 largest initial scale (i.e. $8W$) being too large. We therefore
 provided a parameter, henceforth referred by $largestscales$, that
 allows users to overwrite the default when needed. This prevents
 \WAsp\ from inadvertently fitting large scale negative side-lobes in
 extreme cases, like the data set used for images in
 Section~\ref{Sec:jet}. This also improves the imaging performance for
 datasets that have partially-measured structure on the largest
 spatial scales, where $8W$ may be too conservative to reconstruct a
 good model, like the dataset in Sec.~\ref{Sec:papersky}.

 When the $largestscale$ is set, Eq.~\ref{Eq:init_scales}  is used to 
 define the initial scale sizes ranging from 0 to the $largestscale$ size. 

 \begin{equation}
 \sigma_i = 2^i W, \quad \text{for } i \geq 0 \text{ and } 2^i W \leq \text{largestscale}
 \label{Eq:init_scales}
 \end{equation}

\subsubsection{Update model and residual image using the latest aspen}
\label{Sec:LatestAspen}
The \Asp-Clean is optimized for narrow-band imaging and uses a
heuristic approach to update the model and the residual images with
significant Aspen (Step 2(c) of Section~\ref{Sec:asp}). We simplified
this by using only the latest Aspen of each iterations to update the
model and the residual images. That is,
\begin{eqnarray}
\vec{I}^M_k &=& \vec{I}^M_{k-1} + P_k(\vec{p}) \label{Eq:updateM}\\
\vec{I}^R_k &=& \vec{I}^R_{k-1} - \tens{B}P_k(\vec{p}) \label{Eq:updateR}
\end{eqnarray}
where $P_k(\vec{p})$ is the latest Aspen.  Therefore, in every
iteration there is only one Aspen in the active set for optimization
and there is no need to maintain and later optimize a permanent
list. The \Asp-Clean algorithm is to discover a basis set adopted to
the structure in the image. \BF{This change keeps only the latest
  Aspen in the active set which effectively implies an approximation
  where the covariance matrix is assumed to be a diagonal matrix. This
  may lead to needing more Aspen compared to an algorithm that relaxes
  this assumption, but it significantly reduces the overall runtime cost.
  While a more detailed analysis to understand the limits of the
  parameter space beyond which this approach may breakdown is part of
  the future work, we offer here a few intuitive thoughts as a guide.
  First, one can think of this approach as an ``adaptive,
  scale-sensitive'' version of classical {\it scale-insensitive}
  algorithms, where pixels in the image are treated as independent, even
  when reconstructing diffused emission. With algorithms like the
  H\"{o}gbom Clean \citep{Hogbom_Clean} (and its many variants), a
  large number of components are required to reconstruct diffused
  emission leaving large scale residuals.  Similarly, even though a
  relatively larger number of components (Aspen) may be required with
  this approach, due to the {\it scale-sensitive} nature of the
  components with a wider range of allowed scales, one can expect to
  go relatively deeper, especially with diffused emission, before
  hitting saturation (insignificant changes with iterations), or even
  divergence.  Second, we note that the coupling between Aspen may
  become important at relative high dynamic range imaging of complex
  spatio-frequency structures.  Periodic joint optimization of a subset
  of coupled Aspen may optimize both runtime and imaging performance in this
  limit (e.g. the active-set in \cite{Asp_Clean}; also see 
  \cite{Mueller-ClusterClean}).  Finally, we note that such an
  approximation is also standard in many optimization algorithms
  applied to real-life problems \citep{numericalrecipes3}. Existing
  theoretical framework to analyze the limits of such approximations
  may therefore be directly applicable for understanding the limits of
  our approach as well.  Current tests show that using this approach,
the runtime is reduced by 3x-20x compared to the original approach
without degrading the imaging performance.}

In the wide-band configuration, instead of computing the principal
solution for all pixels separately for all scales, and finding the
best component as described in Step 2 of the MS-MFS algorithm in
Section~\ref{Sec:diffmtmfs}, \WAsp\ first performs a modified Step 2
of the \Asp-Clean algorithm in Section~\ref{Sec:asp}.  That is,
\WAsp\ first smooths {\it only} the first-order Taylor coefficient residual image
by the initial scales to find the optimal scale in Step 2(a) of
Section~\ref{Sec:wasp_impl_wide}. This optimal scale is then used as an
initial guess and further optimized in Step 2(b). This optimized Aspen 
is then used to update the model and residual images for all 
Taylor terms (Step 2(c) and Step 2(d) of Section~\ref{Sec:wasp_impl_wide}).

\subsubsection{Fused deconvolution}
\label{Sec:Fused}
The runtime cost of the \Asp\-Clean algorithm is dominated by the numerical
optimization to find the optimal set of scales at each iteration.  As
the iterations proceed, the emissions left in the residual image is
progressively at smaller scales, which is most efficiently and
appropriately modeled by delta function.  This is indeed what was
observed when applying any of the scale-sensitive algorithms, including
the \Asp-Clean algorithm.  Improvement in the runtime performance is
therefore expected with a robust heuristic to detect this state and
switch to {\it scale-insensitive} algorithms (like the H\"{o}gbom
CLEAN) algorithm) which also have significantly lower computational
complexity.

Therefore, to further reduce the \WAsp\ runtime, the following
heuristic algorithm is used to automatically switch \WAsp\ to 
scale-insensitive mode of iterations when either of the
three criteria is met.

\begin{enumerate}
\item The peak residual is smaller than a user-defined parameter, $fusedthreshold$.   
\item The optimized amplitude of the Aspen is smaller than 
  $\epsilon_0 \, fusedthreshold$ and the difference between the current and the 
previous peak residual is less than $\epsilon_1$ for three consecutive
iterations. The default values used in our implementation (determined
empirically) are $\epsilon_0=5e^{-4}$ and $\epsilon_1=1e^{-4}$.
\item In ten consecutive iterations, more than five iterations pick 0 
scales (See \cite{FusedAsp}).
\end{enumerate}

Instead of triggering the H\"{o}gbom CLEAN algorithm for the
scale-insensitive algorithm, we configure \WAsp\ to use only the
0 scale for $N_{H}$ iterations.  This in our implementation achieves
the same result without increasing the software complexity.  When
this is triggered, it picks $N_{H}$ number of zero-scale
components before potentially switching back to \texttt{WAsp}.

In \citep{FusedAsp}, $N_{H}$ is defined in Eq.~\ref{Eq:NumHog}.
\begin{equation}
\label{Eq:NumHog}
N_H = \text{ceil}(100 + 50(e^{0.05T_{n}} - 1))
\end{equation}
where $T_{n}$ is the number of times the zero-scale algorithm is
triggered. The specific form of Eq.~\ref{Eq:NumHog} is less important
and other forms are admissible to save computational time.  Therefore,
we further simplify Eq.~\ref{Eq:NumHog} with $N_H = 51$, or $N_H = 510$
if the difference between the root mean square of the current residual
and that of the initial residual at the beginning of the minor cycle
is less than $50\%$. This condition indicates that the residual has
become more noise-like.  We compared the imaging performance of
\WAsp\ using our simplified $N_H$ and the Eq.~\ref{Eq:NumHog}. The
simplified $N_H$ improves computational efficiency further without
degrading imaging performance.

\begin{figure*}
\centering
\setlength{\fboxsep}{0pt}
\fbox{\includegraphics[width=10cm]{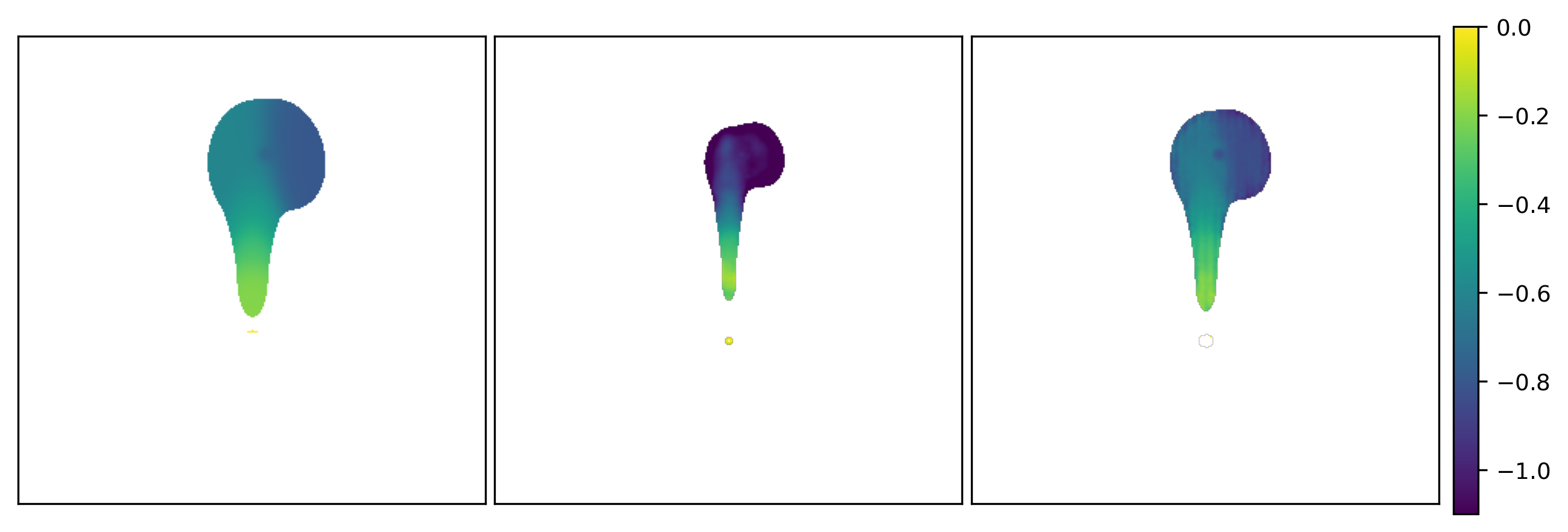}}
\setlength{\fboxsep}{0pt}
\fbox{\includegraphics[width=10cm]{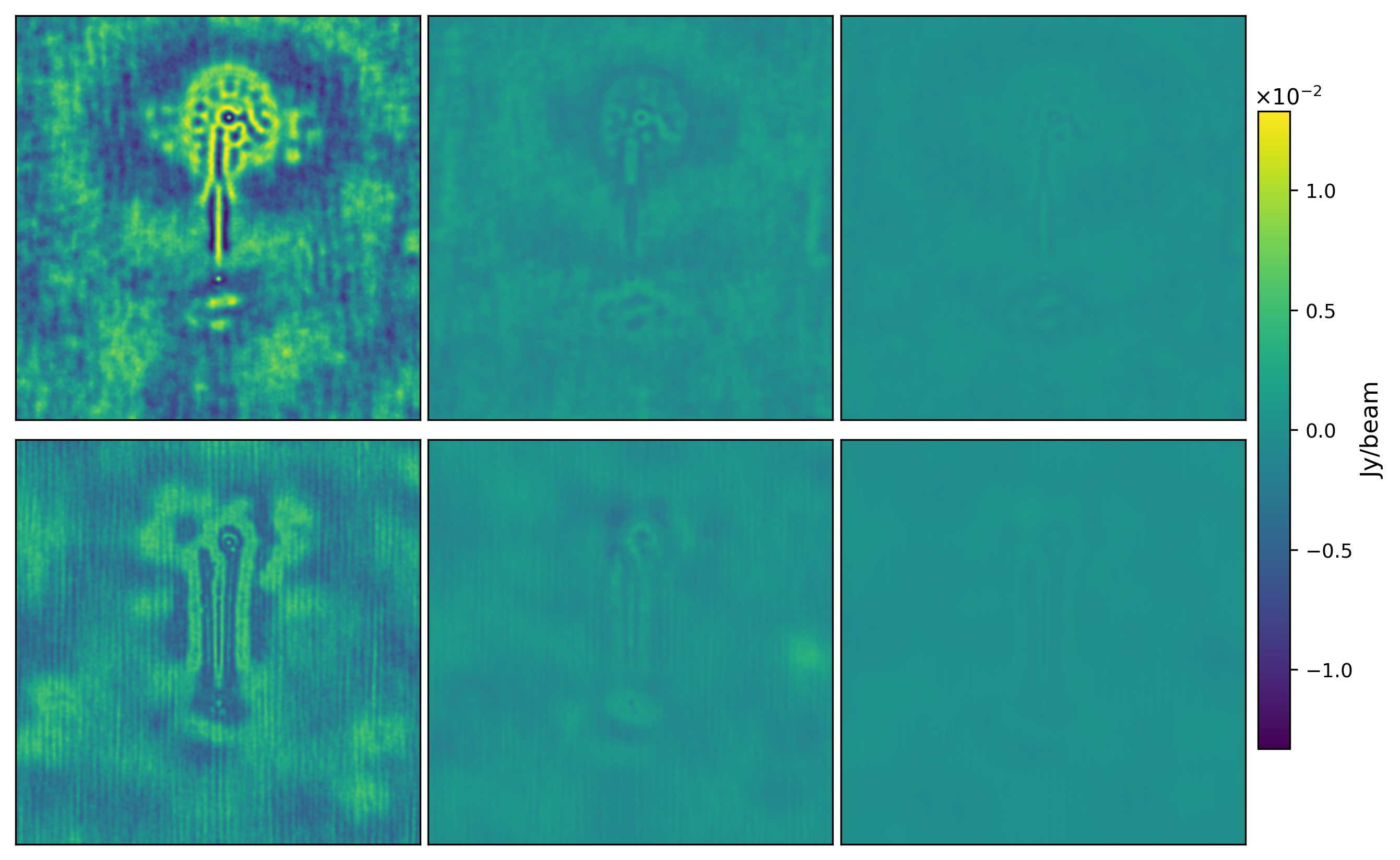}}

\caption{\small Top panel: Spectral index comparison between the truth
  (left), MS-MFS \BF{with {\tt gain=0.1}} (middle), and
  \WAsp\ \BF{with {\tt gain=0.4}} (right) on the jet dataset,
  illustrating the wide-band imaging test of scale-sensitive modeling
  and spectral index reconstruction accuracy across mixed compact and
  extended structures.
  Bottom panel: Residual image comparison of the first three Taylor terms
  between the MS-MFS (top) and \WAsp\ (bottom) on the jet dataset,
  demonstrating the improved wide-band spectral modeling and more
  noise-like residual structure achieved by \WAsp. \BF{All images in
    each panel are displayed at a common scale shown in the colorbar
    on the right of each panel.}}
\label{Fig:index-jet}
\end{figure*}

\begin{figure*}[h]
\centering
\setlength{\fboxsep}{0pt}
\fbox{\includegraphics[width=10cm]{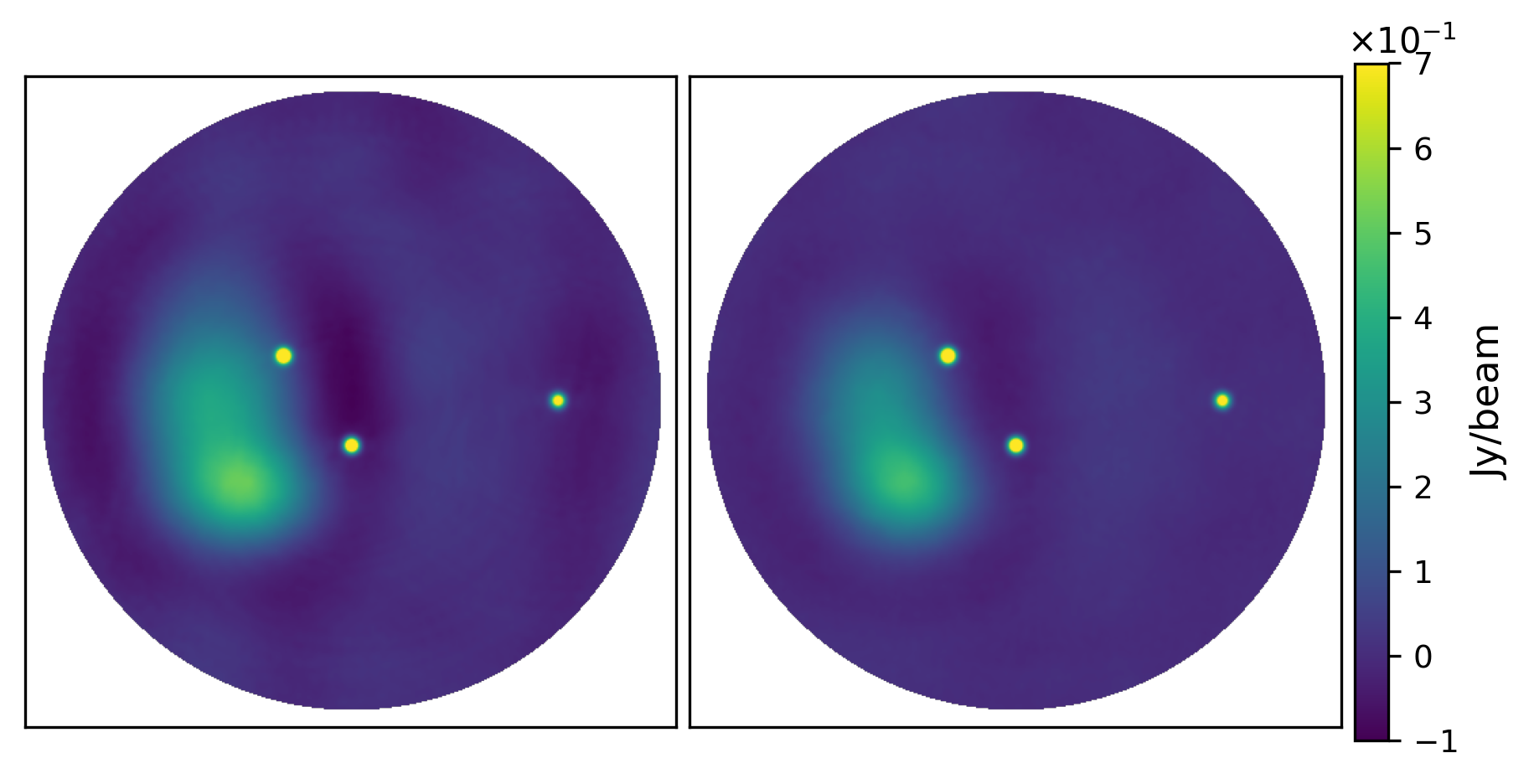}}
\setlength{\fboxsep}{0pt}
\fbox{\includegraphics[width=10cm]{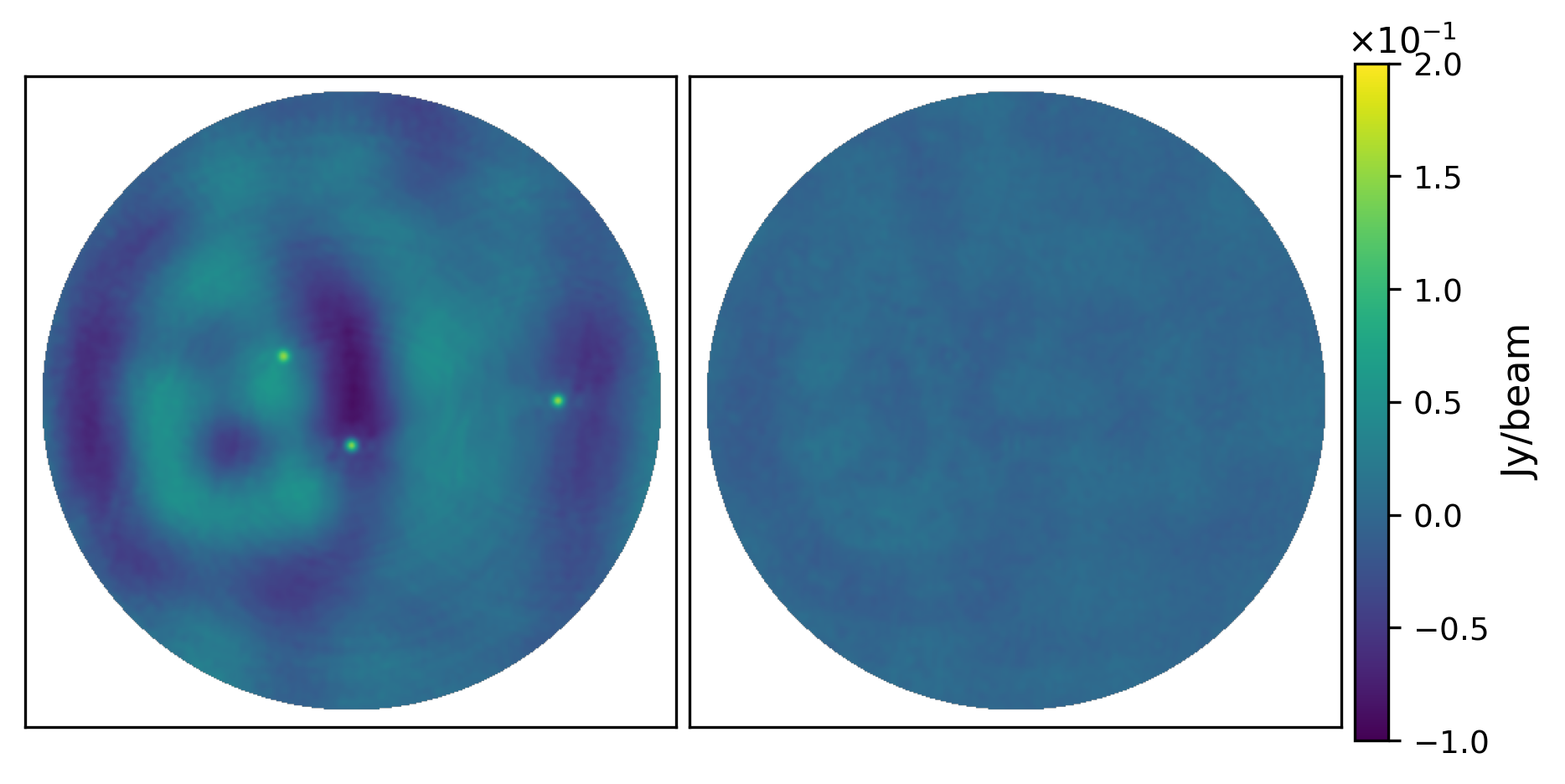}}
\caption{\small Top: Restored image comparison between MS-Clean (left)
  and \WAsp\ (right) on the {\tt papersky} dataset, representing a
  narrow-band imaging test with incomplete large-scale uv coverage to
  assess scale selection behavior.  Bottom: Residual image
  comparison between MS-Clean (left) and \WAsp\ (right) on the
  {\tt papersky} dataset.  }
\label{Fig:image_papersky}
\end{figure*}

\subsection{Relation to MS-MFS} 
\label{Sec:diffmtmfs}
The MS-MFS pre-computes the $N_t N_s \times N_t N_s$ Hessian matrix once at 
the beginning, approximates the Hessian with a block-diagonal structure, 
and stores the inverses of the resulting $N_s$ matrices of size 
$N_t \times N_t$. The pre-computed inverses are then reused throughout 
the iterative process to compute the principal solution and select the 
optimal Taylor coefficients across pixels and scales by the MS-Clean algorithm.  

As described in Section~\ref{Sec:LatestAspen}, the \WAsp\ algorithm
uses the improved \Asp-Clean algorithm instead
(Section~\ref{Sec:diffasp}) to dynamically determine the optimal scale
at the beginning of every minor cycle (Step 2 in
Section~\ref{Sec:wasp_impl_wide}). Thus, the value of $N_s$ in the
\WAsp\ algorithm is always $2$, which indicates the zero scale and the
optimal scale. The Hessian matrices and the inverses are then computed
every cycle at the two scales.  The \WAsp\ algorithm uses the
multi-term multi-frequency imaging part of MS-MFS to solve for the
coefficient matrix $\tens{C}$ as
\begin{equation}
\label{Eq:Coeff}
\tens{C} = \tens{H}^{-1} \ast \vec{I}^D \ast P(\vec{p})
\end{equation}
where $\tens{H}$ is the Hessian matrix, and is given by $\tens{B}
\ast P(\vec{p})\ast \tens{B} \ast P(\vec{p})$ and $\ast$
represents convolution.

It may seem straightforward to optimize $P(\vec{p})$ by minimizing 
$\|\vec{I}^R - \tens{C} \ast \tens{B}\ast P(\vec{p})\|^2$ 
across all Taylor term images, resulting in $N_t$ Aspen. However, 
we find this to be more complex than necessary. \WAsp\ achieves 
significantly better performance by optimizing 
$P(\vec{p})$ {\it only} for the $TT_0$ image. The optimized $P(\vec{p})$ 
is then used in Eq.~\ref{Eq:Coeff} to compute $\tens{C}$, allowing 
the model and residual images for all Taylor terms to be updated 
while preserving most of the numerical benefits of the improved 
algorithm on \(TT_0\) alone.

In the narrow-band configuration, the coefficient matrix $\tens{C}$ 
reduces to a single Taylor term. Alternatively, \WAsp\ employs a 
normalization method (Appendix~\ref{Sec:NormMethod}) to provide 
an initial amplitude estimate, which is then refined by a numerical 
optimization algorithm (Appendix~\ref{Sec:OptLib}) to obtain a best-fit 
value. This combination of normalization and amplitude optimization 
in \WAsp\ can be interpreted as an approximation to solving 
Eq.~\ref{Eq:Coeff}. In practice, it offers better performance 
for narrow-band imaging compared to directly solving Eq.~\ref{Eq:Coeff}. 
For the wide-band configuration, however, Eq.~\ref{Eq:Coeff} is 
solved explicitly to compute $\tens{C}$, providing an effective 
trade-off between algorithmic simplicity and imaging accuracy.

\subsection{Implementation of the \WAsp\ algorithm}
\label{Sec:wasp_impl}
In this section, we describe the \WAsp\ algorithm in its narrow- and wide-band
configurations as implemented in the {\tt LibRA} framework
\citep{AlgoArch, LibRA}.  The Aspen $P(\vec{p})$ is defined using the same functional 
form as in \citep{Asp_Clean} as follows:
\begin{equation}
\label{Eq:2DGaussian}
P = \frac{1}{\sigma\sqrt{2\pi}}e^{-\frac{\left(x-x_0\right)^2+\left(y-y_0\right)^2}{2\sigma^2}}
\end{equation}
where $\vec{p}=\{\sigma ,x_0, y_0\}$, $\sigma$ is the scale size and $(x_0,
y_0)$ is the location of the component. 

\subsubsection{Narrow-band imaging}
\label{Sec:wasp_impl_narrow}
\begin{enumerate}
\item Define a set of initial Aspen as described in
  Section~\ref{Sec:InitScales}.  Let us call this the set $\{P^0_i\}$, 
  where \( i \) ranges from \( 0 \) to \( s \) (the number of initial scales).
\item In each iteration $k$, compute residual images smoothed by the
  set $\{P^0_i\}$.

  \begin{enumerate}
  \item Search for the global peak ($F$) among these 
  smoothed residual images to construct $\BF{P_k(\vec{p})}$ 
  (also see Appendix~\ref{Sec:NormMethod}).
  \item Optimize the $\BF{P_k(\vec{p})}$ (also see Appendix~\ref{Sec:ObjFunc}).
  \item Compute the model image and update the residual image.
  \item Go to Step 2 unless the termination criteria is met.
  \end{enumerate}
\end{enumerate}

\subsubsection{Wide-band imaging}
\label{Sec:wasp_impl_wide}
\begin{enumerate}
\item Define a set of initial Aspen as described in
  Section~\ref{Sec:InitScales}.  Let us call this the set $\{P^0_i\}$, 
  where \( i \) ranges from \( 0 \) to \( s \) (the number of initial scales).
\item In each iteration $k$, compute the first-order Taylor coefficient 
(i.e. $TT_0$) residual image $\vec{I}_R$ smoothed by the set $\{P^0_i\}$.
  \begin{enumerate}
  \item Search for the global peak ($F$) among these smoothed residual
    images to construct $\BF{P_k(\vec{p})}$ (also see Appendix~\ref{Sec:NormMethod}).
  \item Optimize the $\BF{P_k(\vec{p})}$ {\it only} for the $TT_0$ image. 
  (also see Appendix~\ref{Sec:ObjFunc}). 
  \item Update model images by accumulating a set of $N_t$ model
    images, each containing one $\delta$-function (denoted 
    $\vec{I}_{p,k}^{shp}$, where $p$ represents the scale of the optimized 
    $\BF{P_k(\vec{p})}$) that marks the location of the center of the 
    $\BF{P_k(\vec{p})}$. The amplitudes of 
    these $N_t$ $\delta$-functions are the Taylor coefficients that model 
    the spectrum of the integrated flux of the $\BF{P_k(\vec{p})}$. Let these 
    $N_t$ Taylor-coefficient images be denoted as 
    $\vec{I}_{p,q,k}^{M}$; $q \in [0, N_t]$. The set of $N_t$ model images 
    is accumulated as follows. 

    \begin{equation}
    \vec{I}_q^{M} = \vec{I}_{q}^{M} + g(\vec{I}_{p,q,k}^{M} * \vec{I}_{p,k}^{shp}) \forall q \in [0, N_t]
    \label{Eq:updateM_msmfs}
    \end{equation}

  \item Update the RHS residual images by evaluating and subtracting
    out the entire LHS of the normal equations.
  \item Repeat from Step 2 until the minor-cycle flux limit is
    reached.
  \end{enumerate}
\end{enumerate}

\section{\WAsp\ Imaging Results}
\label{Sec:results}

\subsection{EVLA Simulation}
\label{Sec:jet}
To evaluate the effectiveness of the \WAsp\ algorithm for wide-band
imaging, we use a simulation of a jet and lobe-like structure to
represent a realistic scenario (Fig.~\ref{Fig:index-jet}, top
left). This simulation includes thin jets as well as a combination of
compact and extended structures with complex spectral indices. The
colors in Fig.~\ref{Fig:index-jet} (top panel) illustrate the
variations in the spectral index. The long, narrow portion of the
source is referred to as the jet, where the spectral index varies
along its length, ranging from $-0.2$ to $-0.8$. At the top of the
jet, we simulate large-scale emission to represent a hot-spot, where
the spectral index varies between approximately $-0.5$ and $-0.8$,
superimposed with a compact source exhibiting a spectral index of
$-1.0$.

The dataset consists of 5 channels, covering the frequency range of 1 -- 2~GHz, \BF{simulated for the
VLA in the D configuration}. The top panel of Fig.~\ref{Fig:index-jet} compares the results of the 
MS-MFS and \WAsp\ algorithms. The middle image presents the reconstruction obtained 
using the MS-MFS algorithm, while the image on the right shows the reconstruction 
using the \WAsp\ algorithm, where the reconstructed spectral index is significantly 
closer to the true values. The bottom panel of the figure displays the residual images 
of the first, second, and third-order Taylor coefficients for both the MS-MFS (top) 
and \WAsp\ (bottom) algorithms. Notably, the residual images for the \WAsp\ algorithm 
exhibit more noise-like characteristics compared to the MS-MFS algorithm.

\subsection{EVLA Simulation for Non-Standard Conditions}
\label{Sec:papersky}
To evaluate the methodology for determining the initial scale sizes 
(Section~\ref{Sec:InitScales}), the narrow-band \WAsp\ algorithm was tested using a 
simulated dataset, referred to as the {\tt papersky} dataset. This dataset was 
designed to feature partially measured structures at the largest spatial scales (at 
and beyond the uv-hole). The lowest channel fails to capture these large-scale 
structures, making strict reconstruction impossible; however, the second and third 
channels provide sufficient coverage. This under-constrained scenario necessitates the 
use of a mask with the MS-Clean algorithm in CASA (\cite{CASA_Doc}) to prevent 
divergence in the reconstruction.

The top panel of Fig.~\ref{Fig:image_papersky} compares the restored images obtained 
with the MS-Clean and \WAsp\ algorithms. The \WAsp\ algorithm does not require a mask 
to converge and demonstrates superior imaging performance. The bottom panel of this 
figure shows that the residual images for \WAsp\ are more noise-like compared to those 
of MS-Clean. In the \WAsp\ results, the largest scale was set to 70 
(i.e. $largestscale=70$), although using the default largest scale, $8W$, also yields 
comparable results.

\subsection{VLA Observations of Cygnus-A}
\label{Sec:cygA}
To demonstrate the effectiveness of the \WAsp\ algorithm on real data
in the narrow-band configuration, it was applied to high dynamic-range 
VLA observations of the source
Cygnus A. The data was taken as part of a multi band observational
campaign on Cyg A, and is used here with the permission of the
investigators (Perley, private communication). The particular subset
of data used for this test consisted of a single frequency channel of
2MHz channel width centered on 2.052 GHz. In the test setup, imaging 
was performed with $N_t = 1$, at a resolution of 0.2 arcsec and with 
an image size of 2048 $\times$ 2048 pixels.

\begin{figure*}
\centering
\setlength{\fboxsep}{0pt}
\fbox{\includegraphics[width=10cm]{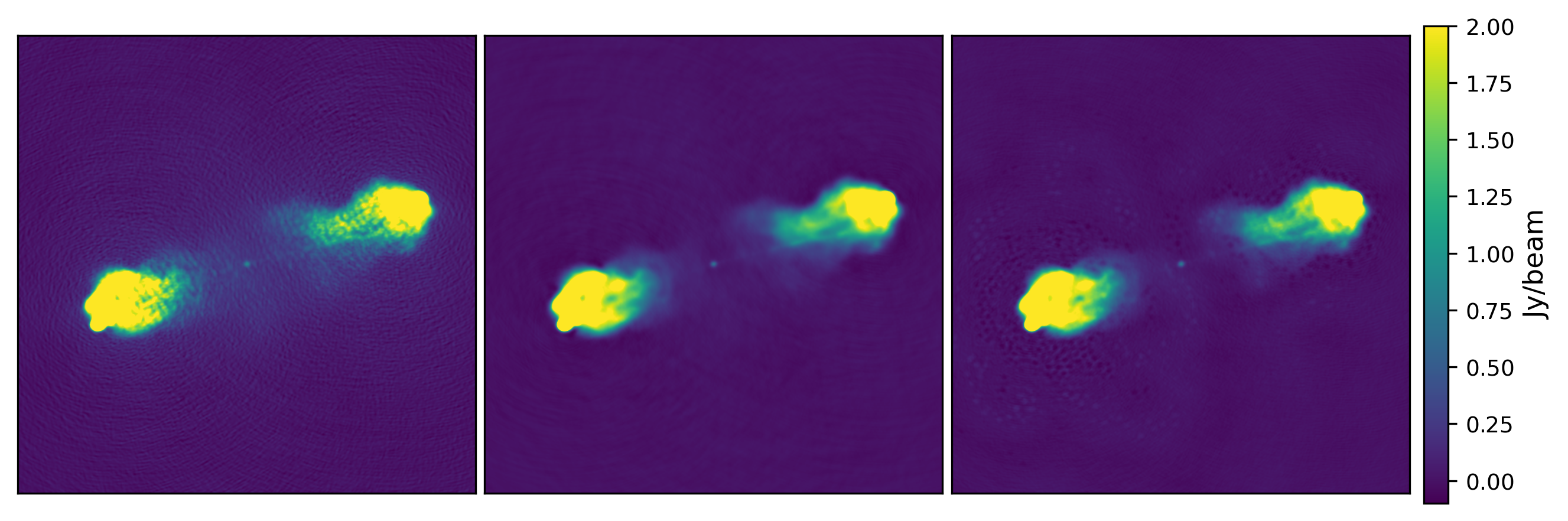}}
\setlength{\fboxsep}{0pt}
\fbox{\includegraphics[width=10cm]{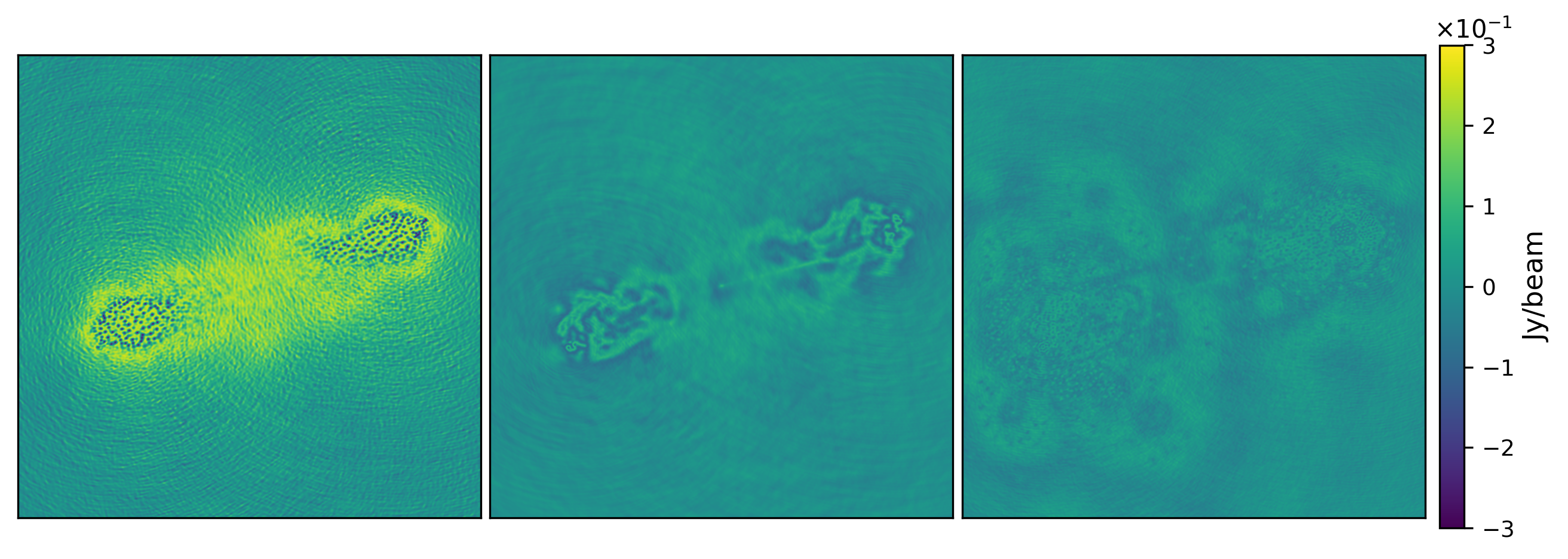}}
\caption{\small Top: Restored image comparison between H\"{o}gbom
  CLEAN (left), MS-Clean (middle) and \WAsp\ (right) on the Cyg A
  dataset, representing a real-data narrow-band imaging test of scale
  modeling.  Bottom: Residual image comparison between H\"{o}gbom
  CLEAN (left), MS-Clean (middle) and \WAsp\ (right) on the Cyg A
  dataset, highlighting the improved convergence behavior and more
  noise-like residual structure obtained with \WAsp. \BF{All images in
    each row are displayed at a common scale shown in the colorbar
    on the right of each row.} }
\label{Fig:image_CygA}
\end{figure*}

The top panel of Fig.~\ref{Fig:image_CygA} compares the restored images for the
H\"{o}gbom CLEAN, MS-Clean and \texttt{WAsp}. MS-Clean was run with five scales
of sizes 0, 10, 20, 30 and 60 pixels. Even under the narrow-band condition, 
\WAsp\ shows robust performance, delivering better imaging without 
requiring a highly tuned mask or a highly fine-tuned list of scales. The bottom 
panel of the figure shows that the \WAsp\ residual image is more noise-like 
than those from H\"{o}gbom CLEAN and MS-Clean.

\begin{figure*}
\centering
\setlength{\fboxsep}{0pt}
\fbox{\includegraphics[width=10cm]{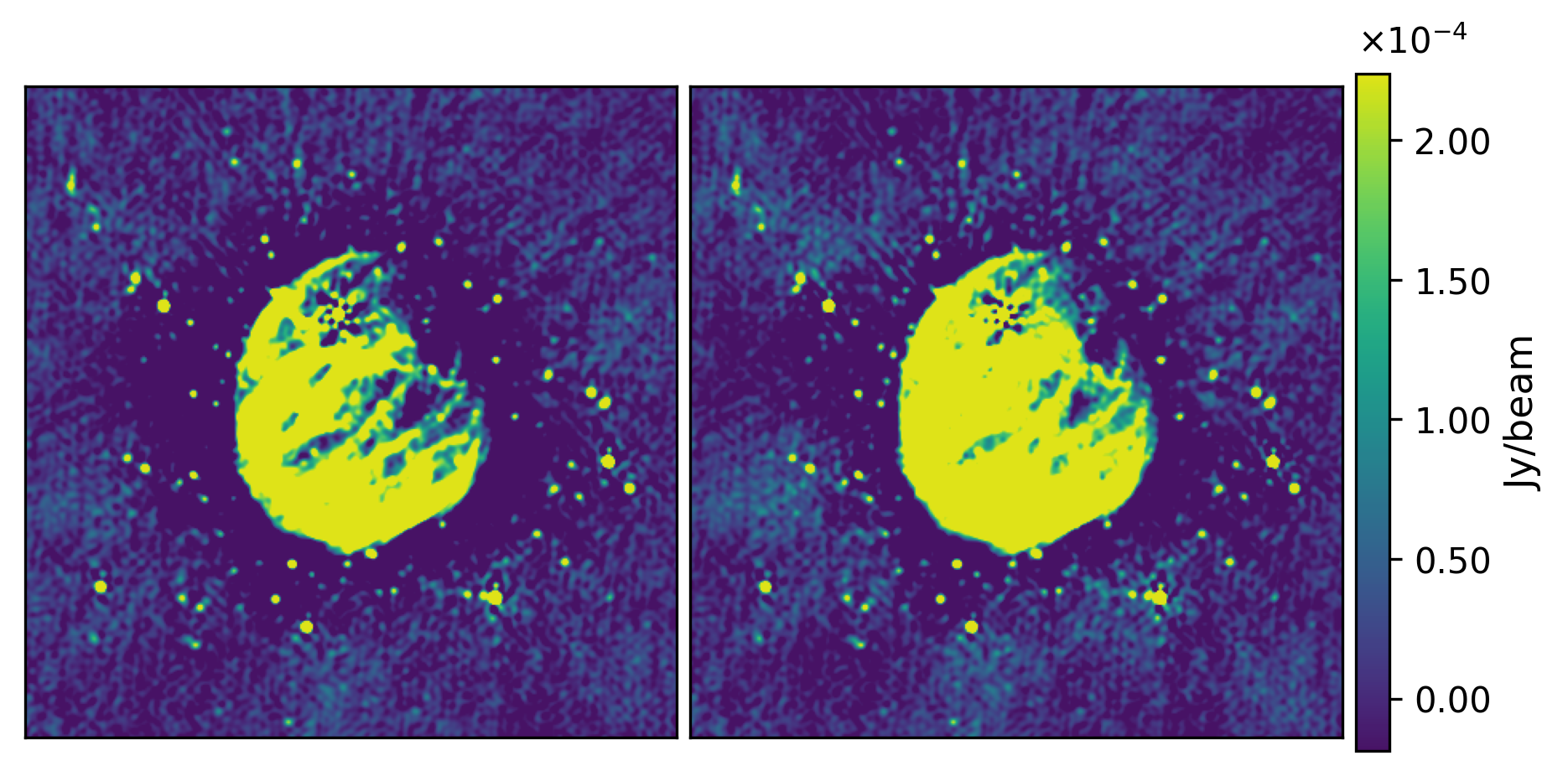}}
\setlength{\fboxsep}{0pt}
\fbox{\includegraphics[width=10cm]{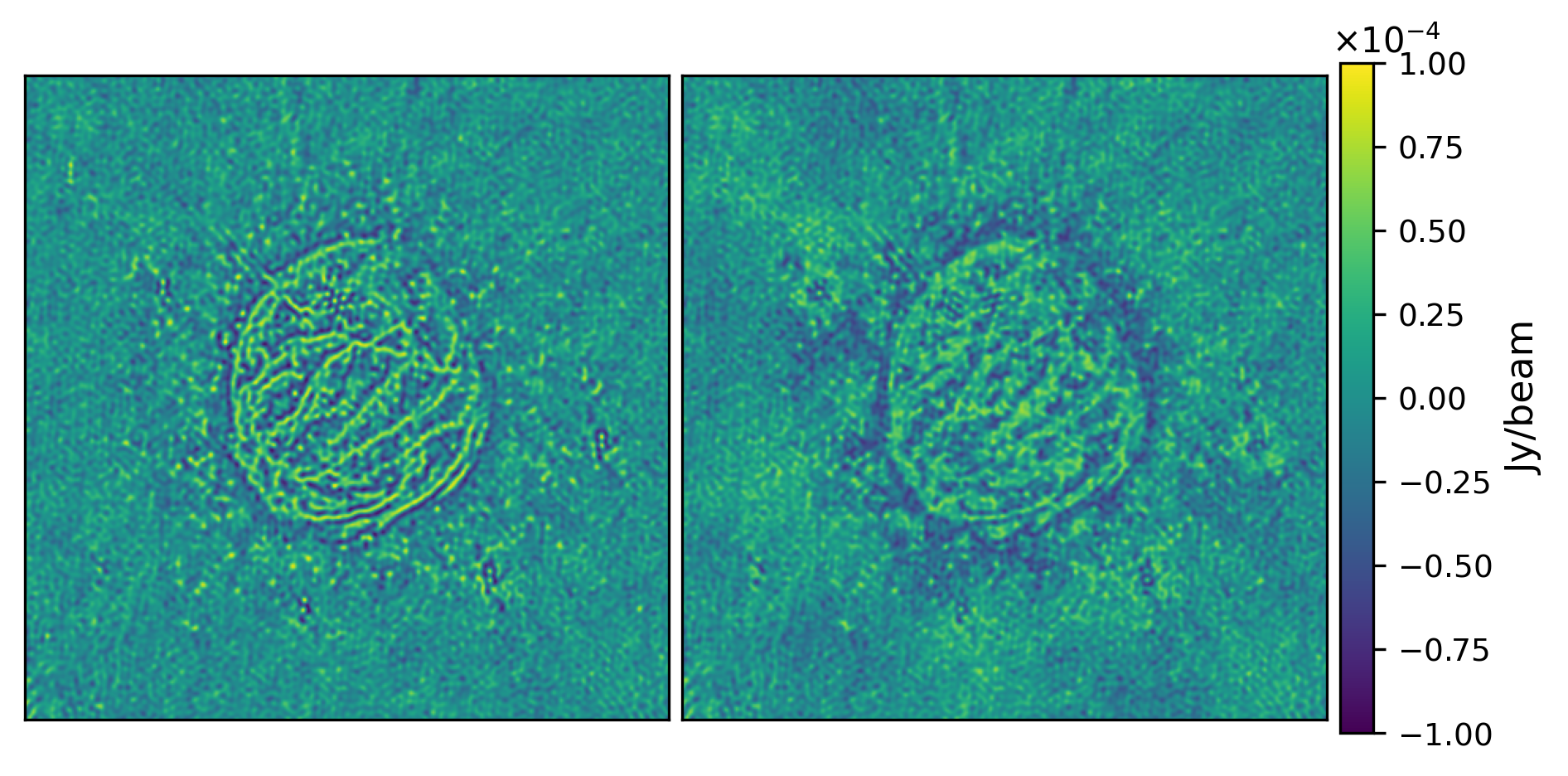}}
\setlength{\fboxsep}{0pt}
\fbox{\includegraphics[width=10cm]{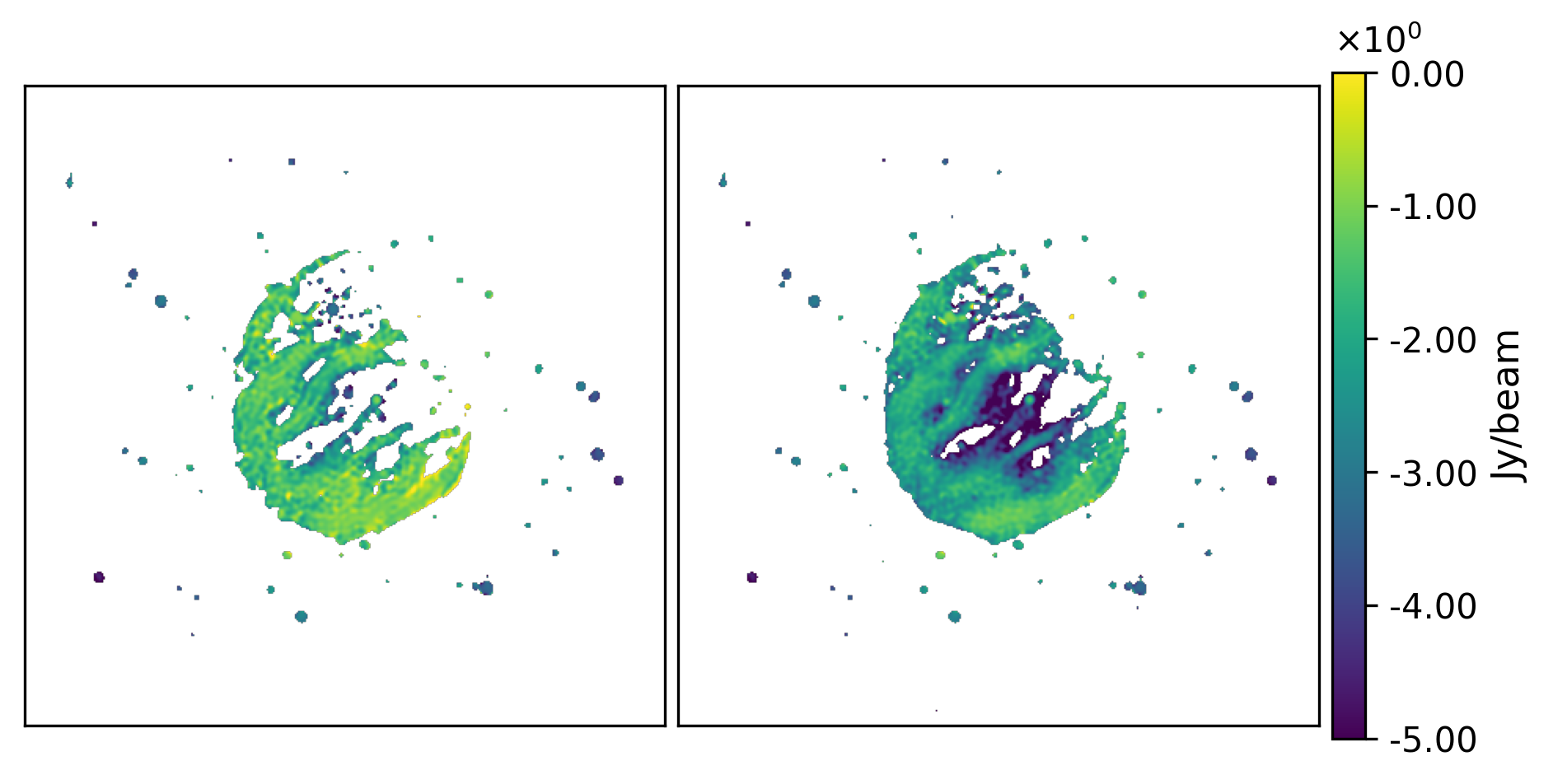}}
\caption{\small Top: Restored image comparison between the MS-MFS (left)
  and \WAsp\ (right) on the G055.7+3.4 dataset, representing a real-data
  wide-band imaging test of scale modeling and spectral index recovery.
 Middle: Residual image comparison between the MS-MFS (left)
  and \WAsp\ (right) on the G055.7+3.4 dataset.
 Bottom: Spectral index comparison between the MS-MFS (left)
  and \WAsp\ (right) on the G055.7+3.4 dataset. \BF{All images in
    each row are displayed at a common scale shown in the colorbar
    on the right of each row.}}
\label{Fig:index-G55}
\end{figure*}

\subsection{Multi-Frequency Observations of a Supernova Remnant G055.7+3.4}
\label{Sec:G55}
The G055.7+3.4 is a supernova remnant with a pulsar within it and is
an extended source with many angular scales. To demonstrate the
effectiveness of the \WAsp\ algorithm, we applied it on wide-band real
data (\BF{data taken in the D-array configuration of
  the VLA at L-band with 512 channels each 2~MHz-wide, covering the
  frequency range of 1 -- 2.03 GHz with $\sim 6$ hours of on-source time
  with an integration time of 2 seconds per sample adding up to a
  data volume of 25~Gigabytes; for more
  details see \cite{Bhatnagar_G55}}). Fig.~\ref{Fig:index-G55} presents
the imaging results where we compare the restored images, residual
images and the spectral index map of G055.7+3.4 obtained with the
MS-MFS (left) and \WAsp\ (right) algorithms.  The MS-MFS
reconstruction used five spatial scales (0, 6, 10, 30, and 60 pixels)
and three Taylor coefficients in the spectral model.

In the restored images (top panel), the negative bowl is less deep 
in the \WAsp\ result, indicating a better fit to 
the larger scales. The middle panel shows that the \WAsp\ 
residual image exhibits more noise-like characteristics than that of MS-MFS.

The spectral index maps (bottom panel) require more careful interpretation. 
No wide-band primary beam correction was applied to the measurement set; consequently, 
the derived spectral index steepens progressively away from the field center.
In addition, the missing short-spacing (uv-hole) information makes the reconstruction 
of the spectrum of the largest scales unconstrained. Since \WAsp\ employs the same 
spectral model as MS-MFS, it does not compensate for this limitation. Therefore, 
the evaluation of spectral performance focuses on the finer-scale structures 
where the signal-to-noise ratio is high. In these regions, the \WAsp\ spectral 
index map is smoother and more closely follows the filamentous morphology 
apparent in the Stokes-I image.

\begin{figure}
    \centering
    \setlength{\fboxsep}{0pt}
    \fbox{\includegraphics[width=0.4\textwidth]{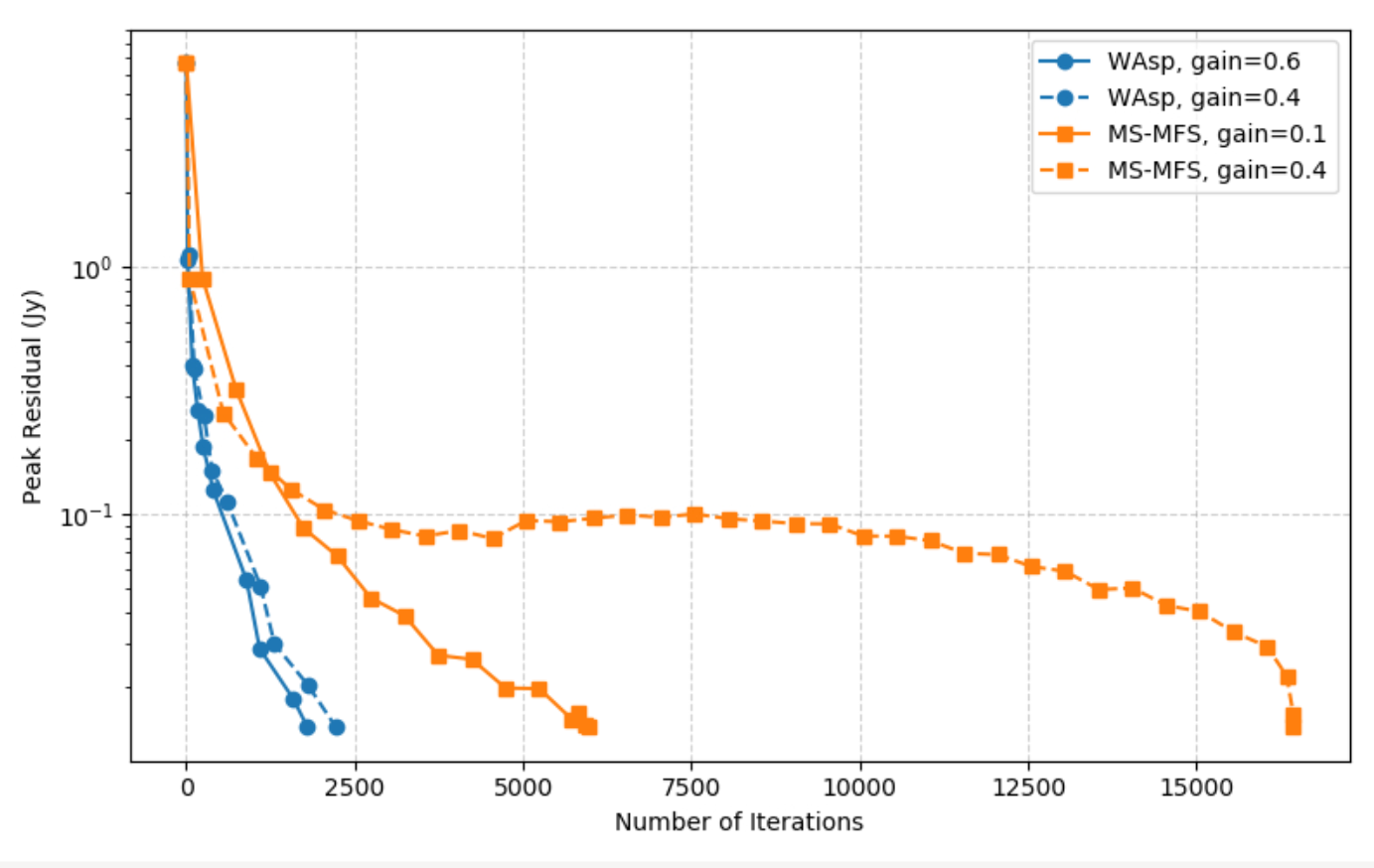}}
    \caption{Comparison of the convergence progress between the
      \WAsp\ and MS-MFS algorithms \BF{with the jet dataset},
      illustrating their relative rates of approaching the final
      solution.  \BF{Markers along each curve indicate where the
        calculations for the {\tt UpdateDir} (``major cycle'') step
        were triggered.}}
    \label{Fig:convergence}
\end{figure}

\subsection{Performance Analysis}
\label{Sec:perform}
Building upon the superior imaging performance demonstrated in the previous sections, 
we now examine the runtime characteristics of \WAsp\ in comparison with MS-MFS and the 
original \Asp-Clean using the jet dataset described in 
Section~\ref{Sec:jet} as a representative wide-scale test case. 

\BF{Figure~\ref{Fig:convergence} compares the rate of
  convergence of the \WAsp\ and MS-MFS algorithms.  The blue curves show
  the convergence profiles for \WAsp\, while the orange curves correspond to
   MS-MFS.  Markers along each curve indicate the iterations where 
   the {\tt UpdateDir} (``major cycle'') step were triggered.}  
The \WAsp\ algorithm achieved convergence more
efficiently while using a substantially higher loop gain (0.6 versus
0.1 for MS-MFS). The lower gain value for MS-MFS was chosen
deliberately, as it required smaller iterative steps to maintain
numerical accuracy. In contrast, \WAsp\ demonstrated sufficient
robustness to operate stably at a much higher gain, achieving superior
spectral index reconstruction even compared to MS-MFS run with the
more conservative gain setting. This highlights that \WAsp\ not only
converges effectively but also tolerates more aggressive updates
without compromising imaging accuracy.

\BF{To assess the impact of loop gain on MS-MFS, we performed
  additional runs on the jet dataset with gain = 0.4. Without
  additional control, the algorithm diverges due to overly large
  iterative updates (data for this is not shown in
  Fig.~\ref{Fig:convergence}). Enforcing smaller effective step sizes
  by triggering more frequent {\tt UpdateDir} evaluations stabilizes
  the solution, but increases the number of time these evaluations are
  triggered from 15 to 35 (as shown in
  Fig.~\ref{Fig:convergence}). Thus, while higher loop gains can be
  used in MS-MFS, they require smaller iterative steps and more
  frequent global updates, leading to significantly slower convergence
  in terms of {\tt UpdateDir} evaluations.}

It is important to emphasize that, due to the nature of the wide-band
imaging problem, the per-iteration runtime is not, by itself, a
meaningful performance metric. The cost per iteration depends strongly
on the structural complexity of the sky brightness distribution. What
ultimately determines practical performance is the total time to
convergence, which is governed primarily by the number of {\tt
  UpdateDir} evaluations. This is clearly illustrated by the jet
dataset, where the runtime is not dominated by the {\tt UpdateDir}
step. Although the per-iteration cost of \WAsp\ is approximately five
times higher than that of MS-MFS \BF{(at gain = 0.1)},
\WAsp\ converges in 9 major cycles, six fewer than MS-MFS, and the
resulting total wall-clock time-to-convergence is comparable.  \BF{In
  contrast, when MS-MFS is operated at gain = 0.4, it requires 35 {\tt
UpdateDir} evaluations, and therefore incurs a substantially
  longer total time-to-convergence compared to \WAsp.}  Thus, a higher
    per-iteration cost does not necessarily translate into longer
    time-to-convergence when convergence is achieved more efficiently.

We also conducted the performance analysis with the real wide-band G055.7+3.4 dataset 
(Section~\ref{Sec:G55}). 
In these runs where imaging conditions are comparable, MS-MFS required 7 major cycles to 
converge with the total wall-clock runtime of 56 minutes, while \WAsp\ reached convergence 
in 5 major cycles with the total wall-clock runtime of 33 minutes. For \WAsp, only about \BF{4 -- 5} 
minutes of the total 33 minutes were spent in minor cycles, indicating that the major cycle 
already dominates the runtime for this dataset. The reduction from 7 to 5 major cycles 
therefore has a significant impact on the total wall-clock time. In other words, 
even a modest reduction in the number of expensive {\tt UpdateDir} evaluations yields 
substantial runtime savings. 

This behavior is consistent with the algorithmic design of \WAsp. By selecting the optimal 
scale at each iteration, the sky model is represented more compactly and accurately, reducing 
the number of global updates required to reach convergence. The primary performance gain 
therefore comes not from faster minor cycles per se, but from requiring fewer major cycles.

For next-generation instruments such as ngVLA and SKA, the major cycle is expected to 
dominate the imaging cost even more strongly. Current projections indicate that these 
steps may require thousands of GPUs operating continuously. In that regime, the 
appropriate performance metric is the total number of major cycle evaluations required 
to converge, rather than the runtime of an individual iteration. A full characterization with 
ngVLA-scale simulations is ongoing, but the present results indicate that \WAsp\ reduces the 
number of {\tt UpdateDir} evaluations and thereby directly targets the dominant computational 
bottleneck. 

For narrow-band imaging (i.e. $N_t = 1$), \WAsp\ demonstrates even
greater improvements, running approximately $20\times$ faster than the
original \Asp-Clean while maintaining similar imaging
fidelity\footnote{these speed improvements were incorporated into the
  CASA Asp implementation since v6.6.3.}.

Finally, the runtime memory footprint of \WAsp\ is lower than that of MS-MFS because it 
operates on a single optimal scale per iteration rather than maintaining and updating a full 
list of predefined scales. Overall, these results suggest that \WAsp\ provides a favorable 
balance between convergence robustness and computational efficiency, and is well positioned 
for regimes in which major-cycle cost governs overall imaging performance.

\section{Conclusion}
\label{Sec:conclusion}
In this paper, we introduce the \WAsp\ Clean algorithm (Wide-band \Asp-Clean) and 
demonstrate its improvement in the imaging performance of the MS-MFS framework 
for wide-band image reconstruction. This enhancement is achieved by replacing the 
MS-Clean algorithm, which models the sky brightness distribution, with a modified 
\Asp-Clean algorithm, while keeping the underlying multi-term imaging framework 
intact. The \Asp-Clean algorithm identifies the optimal scale of emission at each 
iteration, alleviating the need to reconstruct complex emission using a fixed set of 
scales. As a result, the combination of extended and compact emissions is modeled more 
accurately.

The significant improvement in wide-band imaging reconstruction facilitated by the 
\Asp-Clean algorithm can be understood as follows. The first term in the multi-term 
approach represents the total power image (the Stokes-I image), which models the 
frequency dependence of the brightness distribution. For astrophysical applications, 
this frequency dependence is typically modeled using the spectral index parameter (and 
its variation with frequency). Spectral index mapping involves both the first and 
higher-order terms in the multi-term expansion. Since the first term is dominant, 
small errors in it can translate into substantial relative errors in the spectral 
index maps. Thus, an algorithm that reduces reconstruction errors in the Stokes-I map 
is expected to lead to significant improvements in the spectral index maps. This is 
corroborated by our experiments, which include both carefully simulated data and 
real-world wide-band data.

We evaluated the performance of our algorithm with two simulations designed to 
stress-test the imaging capabilities. The first simulation (Section~\ref{Sec:jet}) 
involves emission at multiple scales, with spectral index variations occurring at 
multiple scales as well. A comparison of the results from the MS-MFS algorithm 
and the \WAsp\ algorithm against the ground truth shows that the \WAsp\ algorithm 
provides substantial improvement in imaging performance, particularly in spectral 
index reconstruction, as expected theoretically. The second simulation, referred to as 
the "{\tt papersky} simulation" (Sec.~\ref{Sec:papersky}), involves compact emission 
superimposed on complex large-scale emission that is under-sampled by the telescope’s 
uv-coverage. This scenario is commonly encountered in practice, and the \WAsp\ 
algorithm demonstrates improvements in imaging performance in this context. Similar 
performance gains compared to both H\"{o}gbom- and MS-Clean algorithms were observed 
in the CygA simulation (Sec.~\ref{Sec:cygA}), which features a more realistic 
brightness distribution.

Finally, we applied the \WAsp\ algorithm to wide-band data of the G55 supernova 
remnant (Sec.~\ref{Sec:G55}), a field with brightness distributed throughout the 
telescope’s field-of-view, with compact sources superimposed. Both the MS-MFS and 
\WAsp\ algorithms accurately reconstructed the spectral index of a compact pulsar 
with a well-established spectrum. Although the largest spatial scales remain 
unconstrained by the uv-coverage and therefore do not yield 
global spectral index improvements, \WAsp\ produces a smoother and structurally 
more consistent spectral index distribution on well-constrained scales.

In addition to imaging performance, it is important to consider the
overall computational cost when using the \WAsp\ algorithm in the
algorithm architecture. The overall computational cost of iterative
optimization involving calculations for the derivative/''major cycle'' and subsequent
model update/''minor cycle'' (the {\tt UpdateDir} and {\tt ModelUpdate} components of
the Algorithms Architecture in \cite{AlgoArch} respectively) is
strongly dependent on the rate of convergence. Combination of
algorithms for these two steps that together achieve higher rate of
convergence, even if this comes at the expense of increased
computational cost of the individual components, ultimately yields a
net improvement in overall runtime efficiency. This consideration
becomes increasingly critical for high-resolution and high-sensitivity
observations from next-generation facilities such as the ngVLA and the
SKA, and is already relevant for many science cases that fully exploit
existing telescopes.  In such regimes, convergence rate and
architectural adaptability to modern computing platforms, featuring
multi-core CPUs, GPUs, and large parallel systems, are more meaningful
performance metrics than the standalone runtime of individual
algorithmic components.

In this work, we demonstrate an improved rate of convergence when
employing the \WAsp\ algorithm for the {\tt ModelUpdate} (``minor
cycle'') step of the algorithm architecture
\BF{(Sec.~\ref{Sec:perform})}. Work is currently in progress for improving
the runtime by adapting this algorithm for multi-core execution points
(CPUs and GPUs) and quantitative characterization of the overall
runtime.  These results\BF{, along with a more comprehensive characterization
  of the imaging and runtime performance of different comparable
  algorithms,} will be presented in future publications. Finally, the
relatively high runtime for \WAsp\ (and other similar algorithms) is
to be fundamentally expected.  They offer better imaging performance
with data from modern telescopes with higher information density (due
to high-resolution and sensitivity, improvements in the electronics
and calibration and imaging algorithms, etc.).  Rate of convergence,
imaging performance and efficient parallelization for architecturally
significant components deployed on multi-core EPs on large parallel
computing platforms are therefore more important criteria to guide
research and development, and more appropriate metrics for evaluating
new imaging algorithms compared to the similar metrics for the
individual components of the algorithm architecture.

The implementation of the \WAsp\ algorithm can also be configured for narrow-band 
imaging, offering runtime performance benefits. We tested this configuration for 
spectral line cube imaging, where wide-band modeling per frequency plane is 
unnecessary. The implementation is now available in the LibRA framework (\cite{LibRA}).

In all of the tests mentioned above, the use of masks to designate regions of 
significant emission was either not required, or only loosely defined masks were 
sufficient (e.g., in the {\tt papersky} and G55 simulations, where large-scale 
emission is under-sampled by the telescope’s uv-coverage). Tightly defined masks or the 
use of computationally expensive (and potentially numerically unstable) algorithms, such as 
automasking, were not necessary. This feature is characteristic of multi-scale 
imaging algorithms in general, and the \Asp-Clean algorithm in particular.

\appendix

\section{Appendix A : Normalization and Optimization}
\label{Appendix:A}

\subsection{Normalization Method for Initial Guess}
\label{Sec:NormMethod}
Normalization in the Step 2(a) of Section~\ref{Sec:asp} is critical for
providing a good initial guess of the scale size and the amplitude of
an Aspen for optimization.  Normalization here does not refer to the
normalization of Gaussian components to have unit area. Normalization
method should be designed to avoid consequent 0 scales at the
beginning, and also the normalized optimal strength (i.e.  initial
guess of the amplitude of an Aspen) cannot be too large. Otherwise,
optimization algorithms will return very large scale size.

The steps below describe the normalization method used in the \WAsp.
\begin{enumerate}
\item Convolved the residual image with the initial scales,
  resulting $\vec{I}^R P_i(\vec{p})$, where \(i\) ranges from \(0\) 
  to \(s\) (the number of initial scales). The initial scales are determined
  based on the methodology described in Section~\ref{Sec:InitScales}.

\item Find the global peak among $\vec{I}^R P_i(\vec{p})$. The global
  peak is denoted, $F$.

\item Normalize $F$ by the normalization method
  described below, and this becomes the initial guess of the amplitude
  of an Aspen for optimization.
\end{enumerate}

The normalization method was developed from the original code presented in 
\cite{Asp_Clean}. It was modified to give the best
imaging results with our simplified approach (Eq.~\ref{Eq:updateM} and Eq.~\ref{Eq:updateR}). The
mathematical details are described below.

\begin{enumerate}
\item Model the initials scales by Eq.~\ref{Eq:2DGaussian} and the residual 
image, $\vec{I}^R$, is convolved with the $i$-th initial scales as followed.
\begin{equation}
\label{Eq:ResConvInitScale}
\vec{I}^R P_i(\vec{p}) =
\vec{I}^R\ast\frac{1}{\sigma_{i}\sqrt{2\pi}}e^{-\frac{\left(x-x_0\right)^2+\left(y-y_0\right)^2}{2\sigma_i^2}}
\end{equation}
for each of the scales, $\sigma_{i}$, except the 0 scale. 

\item Find the global peak among $\vec{I}^R P_i(\vec{p})$.
\begin{equation}
F = \text{global maximum}\left(\vec{I}^R P_i(\vec{p})\right)
\end{equation}

\item The initial guess of the amplitude of an Aspen is calculated as
\begin{equation}
F^\circ = F\sqrt{\frac{d}{2\pi}}
\end{equation}
where $d = \sqrt{\frac{1}{W^2} + \frac{1}{\sigma_{opt}^2}}$,
$W$ is the half-width at half-maximum of the Gaussian fitted to the 
main lobe of the PSF and
$\sigma_{opt}$ is the scale that gives the global peak in Step 2.

\end{enumerate}

\subsection{Objective Function for Aspen Optimization}
\label{Sec:ObjFunc}
The objective function that we want to minimize for optimization is
\begin{equation}
\label{Eq:ObjFunc}
\chi^{2}= \left\| \vec{I}^R - a \cdot(B \ast P(\vec{p})) \right\|^{2}
\end{equation}
where $P(\vec{p}$) is defined in Eq.~\ref{Eq:2DGaussian}, $\vec{I}^R$ is the
residual image and $a$ is the amplitude of the Aspen.

The partial derivatives of $\chi^{2}$ with respect to amplitude and scale are:
\begin{equation}
\label{Eq:DerAmp}
\frac{\partial \chi^{2}}{\partial a} = -2(\vec{I}^R - a \cdot (B \ast P(\vec{p})) \cdot (B \ast P(\vec{p})))
\end{equation}

\begin{equation}
\label{Eq:DerScale}
\frac{\partial \chi^{2}}{\partial \sigma} = -2(\vec{I}^R - a \cdot (B \ast P(\vec{p})) \cdot (a \cdot (B \ast \frac{P(\vec{p})}{\sigma}) \cdot (\frac{\left(x-x_0\right)^2+\left(y-y_0\right)^2}{\sigma^2} - 1))
\end{equation}

It is worth noting that the matrix multiplication operator here,
$\cdot$, should be element by element.

\subsection{Optimization third party libraries}
\label{Sec:OptLib}
The original \Asp-Clean implementation used GNU Scientific Library
(GSL) for optimization in 2004. Since then, many third party libraries
have been developed for better computing performance. All of them have
different API, but the basic requirement is to pass an objective
function for optimization through the API. The derivatives of the
objective function may not be required since some libraries can do the
optimization without derivatives.  A list of optimization libraries
were attempted to be used along with \texttt{WAsp}, and several challenges were
encountered. The final implementation of \WAsp\ uses ALGLIB because it
is not only stable but also computationally more efficient than
GSL. The optimization code is verified by fitting an Aspen to a simple
dirty image (i.e. Gaussian component convolved with a real PSF).

\begin{enumerate}
\item LBFGS++ (\cite{LBFGS++}). It is a header-only C++ library that
  implements the Limited-memory BFGS algorithm (L-BFGS) for
  unconstrained minimization problems, and a modified version of the
  L-BFGS-B algorithm for box-constrained ones. Its API is easy to work
  with.  It requires the Eigen library which is already part of the
  CASA build. However, it is not robust and sometimes returns BFGS
  fitting error due to over-fitting.
\item CppNumericalSolvers (\cite{CppNumSolver}). It is a lightweight
  C++17 library of numerical optimization methods for nonlinear
  functions. Its API is similar to LBFGS++, but returns very different
  results comparing to LBFGS++.
\item GSL (\cite{GSL}). It is a numerical library for C and C++
  programmers and provides wide range of mathematical routines,
  including optimization.  The API is hard to work with and can easily
  cause segmentation fault if not using the API correctly. Debugging
  segmentation fault with {\tt gdb} is not useful and requires special
  GSL API for debugging. A new class was created for passing Aspen
  and data between CASA and GSL. GSL is more stable than the above two
  libraries, and this is probably because GSL provides a restart
  function when an optimization step fails.  GSL provides two methods
  for optimizing without derivatives, {\tt
    gsl\_multimin\_fminimizer\_nmsimplex2} and {\tt
    gsl\_multimin\_fminimizer\_nmsimplex2ran}.  However, the former
  returns very large scale sizes. The latter gives better optimization
  result but its runtime is 3x longer than the optimization method
  with derivatives. Therefore, GSL optimization with derivatives was
  initially used in the \WAsp\ implementation.
 \item ALGLIB (\cite{ALGLIB}). It is a cross-platform numerical
   analysis and data processing library that offers a variety of
   optimization methods.  Performance analysis of the \WAsp\ algorithm
   reveals that the optimization step is a significant bottleneck. To
   address this, the final implementation of \WAsp\ utilizes the more
   efficient ALGLIB library, resulting in a twofold improvement in
   computational performance compared to GSL. However, it is worth
   noting that ALGLIB can exhibit some variability across different
   builds and operating systems. This may necessitate additional steps
   to ensure consistent results, such as adjusting tolerances or
   carefully managing the computational environment.

\end{enumerate}

\subsubsection{Variable scaling for optimization}
\label{Sec:VarScaling}
 In the \WAsp\ algorithm, the parameters of the optimization problem
 are the amplitude and the scale of the Aspen $P(\vec{p})$. These two
 parameters can sometimes exhibit significantly different magnitudes,
 which poses a challenge for convergence. The optimization algorithm
 typically converges only when these parameters are roughly of the
 same order of magnitude, as it employs a single step size for the
 optimization process. To address this scaling issue in \texttt{WAsp},
 we utilize the \texttt{minlbfgssetcscale} function provided by
 ALGLIB. This function informs the optimization algorithm about the
 parameter scales and internally applies transformations to ensure
 that the magnitudes of the parameters are approximately equal,
 thereby facilitating convergence.

\begin{acknowledgements}

  We thank the anonymous referee for useful discussion and feedback.

  The National Radio Astronomy Observatory is a facility of the
  National Science Foundation operated under cooperative agreement by
  Associated Universities, Inc.

\end{acknowledgements}

\software{{\tt LibRA} (ascl:2601.012), {\tt CASA} (ascl:1107.013)}
\facilities{Jansky Very Large Array (VLA) telescope}
\bibliographystyle{aasjournal}
\bibliography{../bhatnagar08}

\begin{thebibliography}{}
\expandafter\ifx\csname natexlab\endcsname\relax\def\natexlab#1{#1}\fi
\providecommand{\url}[1]{\href{#1}{#1}}
\providecommand{\dodoi}[1]{doi:~\href{http://doi.org/#1}{\nolinkurl{#1}}}
\providecommand{\doeprint}[1]{\href{http://ascl.net/#1}{\nolinkurl{http://ascl.net/#1}}}
\providecommand{\doarXiv}[1]{\href{https://arxiv.org/abs/#1}{\nolinkurl{https://arxiv.org/abs/#1}}}

\bibitem[{{Bhatnagar} \& {Cornwell}(2004)}]{Asp_Clean}
{Bhatnagar}, S., \& {Cornwell}, T.~J. 2004, A\&A, 426, 747

\bibitem[{Bhatnagar {et~al.}(2011)Bhatnagar, Rau, Green, \&
  Rupen}]{Bhatnagar_G55}
Bhatnagar, S., Rau, U., Green, D.~A., \& Rupen, M.~P. 2011, The Astrophysical
  Journal Letters, 739, L20, \dodoi{10.1088/2041-8205/739/1/L20}

\bibitem[{Bhatnagar {et~al.}(2025)Bhatnagar, Rau, Hsieh, Kern, \&
  Xue}]{AlgoArch}
Bhatnagar, S., Rau, U., Hsieh, M., Kern, J., \& Xue, R. 2025, The Astronomical
  Journal, 170, 246, \dodoi{10.3847/1538-3881/adfe61}

\bibitem[{Bochkanov(1999)}]{ALGLIB}
Bochkanov, S. 1999, ALGLIB, \url{http://www.alglib.net}

\bibitem[{Cornwell(2008)}]{MS-Clean}
Cornwell, T.~J. 2008, IEEE Journal of Selected Topics in Signal Processing, 2,
  793–801, \dodoi{10.1109/jstsp.2008.2006388}

\bibitem[{{H{\"o}gbom}(1974)}]{Hogbom_Clean}
{H{\"o}gbom}, J.~A. 1974, \aaps, 15, 417

\bibitem[{{Junklewitz, H.} {et~al.}(2015){Junklewitz, H.}, {Bell, M. R.}, \&
  {En\ss{}lin, T.}}]{ResolveClean}
{Junklewitz, H.}, {Bell, M. R.}, \& {En\ss{}lin, T.} 2015, A\&A, 581, A59,
  \dodoi{10.1051/0004-6361/201423465}

\bibitem[{M.~{Galassi} {et~al.}(2009)M.~{Galassi}, {Davies}, {Gough},
  {Jungman}, {Booth}, \& {Rossi}}]{GSL}
M.~{Galassi}, M., {Davies}, J.~T., {Gough}, B., {et~al.} 2009, GNU Scientific
  Library Reference Manual - Third Edition (Network Theory Limited)

\bibitem[{M\"{u}ller \& Bhatnagar(2025)}]{Mueller-ClusterClean}
M\"{u}ller, H., \& Bhatnagar, S. 2025, A\&A, 698, A176,
  \dodoi{10.1051/0004-6361/202553990}

\bibitem[{M\"{u}ller {et~al.}(2026)M\"{u}ller, Hsieh, \&
  Bhatnagar}]{Mueller-ConvexOpt}
M\"{u}ller, H., Hsieh, M., \& Bhatnagar, S. 2026, A\&A, 706, A77,
  \dodoi{10.1051/0004-6361/202555356}

\bibitem[{{Narayan} \& {Nityananda}(1986)}]{MEM_ARAA}
{Narayan}, R., \& {Nityananda}, R. 1986, Annual Reviews of Astronomy and
  Astrophysics, 24, 127

\bibitem[{Offringa \& Smirnov(2017)}]{WSCLEAN}
Offringa, A.~R., \& Smirnov, O. 2017, Monthly Notices of the Royal Astronomical
  Society, 471, 301, \dodoi{10.1093/mnras/stx1547}

\bibitem[{Press {et~al.}(2007)Press, Teukolsky, Vetterling, \&
  Flannery}]{numericalrecipes3}
Press, W.~H., Teukolsky, S.~A., Vetterling, W.~T., \& Flannery, B.~P. 2007,
  Numerical Recipes 3rd Edition: The Art of Scientific Computing (Cambridge
  university press)

\bibitem[{{Qiu}(2015)}]{LBFGS++}
{Qiu}, Y. 2015, LBFGS++.
\newblock \url{https://lbfgspp.statr.me/}

\bibitem[{{Rau} \& {Cornwell}(2011)}]{MSMFS}
{Rau}, U., \& {Cornwell}, T.~J. 2011, \aap, 532, A71,
  \dodoi{10.1051/0004-6361/201117104}

\bibitem[{{The CASA Team, et al.}(2022)}]{CASA_Doc}
{The CASA Team, et al.} 2022, \pasp, 134, 114501,
  \dodoi{10.1088/1538-3873/ac9642}

\bibitem[{{The LibRA Team}(2025)}]{LibRA}
{The LibRA Team}. 2025, LibRA: A library of algorithms for indirect imaging,
  paper in preparation, \dodoi{https://ascl.net/2601.012}

\bibitem[{Wieschollek(2016)}]{CppNumSolver}
Wieschollek, P. 2016, CppOptimizationLibrary,
  \url{https://github.com/PatWie/CppNumericalSolvers}

\bibitem[{{Zhang}(2018)}]{FusedAsp}
{Zhang}, L. 2018, A\&A, 618, A117, \dodoi{10.1051/0004-6361/201833090}

\bibitem[{Zhang {et~al.}(2020)Zhang, Xu, \& Zhang}]{Zhang_2020}
Zhang, L., Xu, L., \& Zhang, M. 2020, Publications of the Astronomical Society
  of the Pacific, 132, 041001, \dodoi{10.1088/1538-3873/ab7345}

\bibitem[{Zhang {et~al.}(2021)Zhang, Mi, Xu, Zhang, Li, Liu, Wang, Xiao, \&
  Wu}]{Zhang_2021}
Zhang, L., Mi, L.-G., Xu, L., {et~al.} 2021, Research in Astronomy and
  Astrophysics, 21, 063, \dodoi{10.1088/1674-4527/21/3/63}

\end{thebibliography}

\end{document}